\documentclass[aps,prl,reprint,floatfix,amsmath,amssymb,twocolumn]{revtex4-2}

\usepackage{epsfig}
\usepackage{graphicx}% Include figure files
\usepackage{dcolumn}% Align table columns on decimal point
\usepackage{bm}% bold math
\usepackage[colorlinks=true,citecolor=blue, dvipdfm]{hyperref}
\usepackage{amssymb}

\begin{document}
%\linenumbers
\title{Complete universal scaling of first-order phase transitions in the two-dimensional Ising model}

\author{Yuxiang Zhang and Fan Zhong}
\email{stszf@mail.sysu.edu.cn}
\affiliation{School of Physics and State Key Laboratory of Optoelectronic Materials and
Technologies, Sun Yat-sen University, Guangzhou 510275, People's Republic of China}

\date{\today}

\begin{abstract}
Phase transitions, as one of the most intriguing phenomena in nature, are divided into first-order phase transitions (FOPTs) and continuous ones in current classification. While the latter shows striking phenomena of scaling and universality, the former has recently also been demonstrated to exhibit scaling and universal behavior within a mesoscopic, coarse-grained Landau-Ginzburg theory. Here we apply this theory to a microscopic model---the paradigmatic Ising model, which undergoes FOPTs between two ordered phases below its critical temperature---and unambiguously demonstrate universal scaling behavior in such FOPTs. These results open the door for extending the theory to other microscopic FOPT systems and experimentally testing them to systematically uncover their scaling and universal behavior.
\end{abstract}
%\pacs{64.60.-i,05.70.Fh.,64.60.My,75.40.Gb}
%\keywords{First-order phase transitions, universal scaling, Langevin equation, renormalization-group theory}

\maketitle

Phase transitions and critical phenomena are among the most intriguing phenomena in nature and society. According to a modern classification~\cite{Fisher67}, phase transitions are divided into first-order and continuous kinds. It is well established that continuous phase transitions are characterized by universal scaling behavior near their critical points~\cite{Stanley,Mask,Cardyb,Justin,Amit}, whereas first-order phase transitions (FOPTs) are typically described by nucleation and growth or spinodal decomposition~\cite{Gunton83,Binder,Binder2,Binder16}. Yet, it has recently been demonstrated that FOPTs also exhibit universal scaling behavior whether in field-driven transitions~\cite{zhong24}, in fixed-field nucleation and growth~\cite{zhong24f}, or even in thermal transitions~\cite{zhong25}. Moreover, the universal scaling is complete. This is because the entire rescaled order parameter curves using universal exponents completely collapse into a single curve over a practically unlimited parameter range in the zero-dimensional models---both with and without noise---without the need for any additional variables. In two dimensions, the rescaled curves collapse onto a single master curve, except in inevitable crossover regions. These results are derived from a mesoscopic, coarse-grained Landau-Ginzburg model for FOPTs via an effective cubic renormalization-group theory~\cite{zhongl05,zhong16}. The essential point is that there exist explicit relationships between the parameters of the coarse-grained model (giving rise to the FOPTs) and its derived effective cubic theory (controlling scaling behavior). However, such explicit relationships are absent in microscopic models. The central challenge in studying universal scaling behavior in real FOPTs then lies in bridging the gap between the theory and real microscopic models.

To this end, we first recapitulate the theory to set the stage~\cite{zhong24}. Consider a free-energy functional~\cite{Mask,Cardyb,Justin,Amit}
%$f_4(\phi)=a_2\phi^2/2+a_4\phi^4/4-H\phi$
\begin{equation}
	F(\phi)=\int d^dx\left\{f_4(\phi)+\frac{1}{2}\left(\nabla\phi\right)^2\right\},\label{gl}
\end{equation}
with
\begin{equation}
	f_4(\phi)=\frac{1}{2}a_2\phi^2+\frac{1}{4}a_4\phi^4-H\phi,\label{f4}
\end{equation}
for an order parameter $\phi$ and its conjugate field $H$ in $d$-dimensional space, where $a_2$ is a reduced temperature and $a_4$ a coupling constant. Dynamics can be studied by the conventional purely dissipative Langevin equation, i.e., Model A dynamics~\cite{Hohenberg}
\begin{eqnarray}
	\frac{\partial\phi}{\partial t}=-\lambda\frac{\delta F}{\delta\phi}+\zeta,\qquad\qquad\qquad\qquad\nonumber\\
\langle\zeta({\bf x},t)\rangle=0,~
\langle\zeta({\bf x},t)\zeta({\bf x}',t')\rangle=2\lambda T\delta(t-t')\delta({\bf x}-{\bf x}'),\nonumber\\\label{lang2d}
\end{eqnarray}
where $\lambda$ is a kinetic coefficient and $\zeta$ a Gaussian white noise with a noise amplitude or temperature $T$.
In a mean-field (MF) approximation in which $\phi$ is spatially uniform, the model exhibits a continuous transition at a critical point $a_2=0$ and $H=0$, but shows field-driven FOPTs between two ordered phases with $\phi_{\rm eq}=\pm\sqrt{-a_2/a_4}=\pm M_{\rm eq}$ at an equilibrium transition point $H=0$ for all $a_2<0$.

In the absence of external perturbation, the FOPTs can only occur beyond the equilibrium transition point $H=0$, where the free-energy cost for homogeneous nucleation diverges. This motivates us to expand $\phi$ and $H$ around a spinodal at $M_{\rm s}$ and $H_{\rm s}$, i.e., $\phi=M_{\rm s}+\varphi$ and $h=H-H_{\rm s}$. Retaining only the leading terms results in an effective cubic theory with~\cite{zhongl05,zhong16}
\begin{eqnarray}
	f_3(\varphi)=\frac{1}{2}\tau\varphi^2+\frac{1}{3}a_3\varphi^3-h\varphi,\qquad~ \label{f3}\\
\tau=a_2+3a_4M_{\rm s}^2,\quad {H}_{\rm s}=a_2M_{\rm s}+a_4M_{\rm s}^3,\label{tauh}	
\end{eqnarray} 
where $a_3=3a_4M_{\rm s}$. $M_{\rm s}$ and $H_{\rm s}$ exactly match the MF spinodal $M_{\rm sMF}=\pm\sqrt{-a_2/3a_4}$, $H_{\rm sMF}=a_2M_{\rm sMF}+a_4M_{\rm sMF}^3=2a_2M_{\rm sMF}/3$, at which the free-energy barrier between the two ordered phases vanishes, allowing the transition in the MF approximation. This indicates that, for the cubic $f_3$, $\tau=0$ and $h=0$ at the MF spinodal point, analogous to $a_2=0$ and $H=0$ at the MF critical point for the quartic $f_4$. Thus, the transition is controlled by the cubic theory $f_3$ near the spinodal, which justifies neglecting higher-order terms in the expansion.

Having established the correct MF theory, we can then systematically renormalize the cubic theory, $F(\phi)$, with $f_3$ replacing $f_4$ in Eq.~(\ref{gl}), to incorporate fluctuation effects. This leads to~\cite{zhongl05,zhong16}
\begin{equation}
	M-M_s=b^{-\beta/\nu}g((\tau+\delta\tau) b^{1/\nu}, (h+\delta h)b^{\beta\delta/\nu}, tb^{-z},a_3^*b^{\epsilon/2}),\label{mheq}
\end{equation}
which is controlled by a nontrivial fixed point $a_3^{*2}=-\epsilon b^{-\epsilon}/6N_d$ below the upper critical dimension $d_c=6$, where $\epsilon=6-d$, $N_d$ is a $d$-dependent constant, $b$ a length scale, $g$ a universal scaling function, $\delta\tau$ and $\delta h$ are cubic fluctuation contributions to $\tau$ and $h$, respectively, and $\beta$, $\delta$, $\nu$, and $z$ are the cubic counterparts of the critical exponents with identical symbols and meaning~\cite{zhongl05,zhong16}. Note that the fixed point is purely imaginary for $\epsilon>0$. Nonetheless, it has been shown that the renormalization drives the system to an unstable point by eliminating modes requiring nucleation; consequently, $a_3$ needs to be analytically continued to imaginary values, allowing the system to automatically converge to the fixed point~\cite{zhong12,zhong16}. Above $d_c$, where $\epsilon<0$, a Gaussian fixed point $a_3^*=0$ takes over. Equation~(\ref{mheq}) remains valid, and MF behavior emerges with $z=2$, $\nu=1/2$, $\beta=1$, and $\delta=2$~\cite{zhongl05,zhong16}, as the system is confined to the effective dimension $d_c$~\cite{Zenged}. This MF theory has been conclusively established to fully catch the complete universal scaling behavior in the zero-dimensional models both with and without noise~\cite{zhong24,zhong24f,zhong25}. In Eq.~(\ref{mheq}), we have neglected corrections to scaling such as deviations from the fixed point~\cite{Wegner,zhong16,Justin,Amit}. 

To reveal the underlying scale invariance controlled by Eq.~(\ref{mheq}) for the FOPTs in the model Eq.~(\ref{gl}), we need to know the relationship between the parameters of Eq.~(\ref{f4}) and Eq.~(\ref{f3}). This is simplified using the method of complete universal scaling, which involves rescaling all parameters properly to achieve curve collapse~\cite{zhong24}. One advantage is that we no longer need to consider fluctuation contributions $\delta\tau$ and $\delta h$ from the $f_3$ theory for $d<d_c$ in Eq.~(\ref{mheq}), since these terms scale as $\tau$ and $h$ themselves. Similarly, we can neglect $M_{\rm s}$, which behaves in the same way as $M$. However, for $d<6$, all $M_{\rm s}$-dependent quantities---notably $a_4$, via $a_3=a_4M_{\rm s}$, and its induced nontrivial fluctuation contributions---possess distinct scaling dimensions from $\tau$ and $h$, requiring a special treatment. Consequently, we derive:
\begin{eqnarray}
	M=b^{-\beta/\nu}g_1(\tau b^{1/\nu}, hb^{\beta\delta/\nu}, tb^{-z},Tb^{[T]},a_3^*b^{\epsilon/2}),\label{mha}\\	
\tau=a_2+3a_4M_{\rm s}^2+\delta a_2,\quad h=H-{H}_{\rm s}+\delta H,~~\label{th}
\end{eqnarray}
where $g_1$ is a scaling function, and we have included the temperature $T$ whose scaling dimension is (denoted by squared brackets) $[T]=z+2\beta/\nu$ originating from the noise~\cite{zhong24}. Equation~(\ref{th}) differs from Eq.~(\ref{tauh}) in the inclusion of $\delta a_2$ and $\delta H$, which encompass all fluctuation contributions from the quartic theory, in contrast to $\delta\tau$ and $\delta h$ derived from the cubic theory. Crucially, Eq.~(\ref{th}) embodies the required special treatment. First, the scaling dimensions of $a_4M_{\rm s}$ and each term in $H_{\rm s}$ (Eq.~(\ref{tauh})) differ from those of $\tau$ and $h$ individually. Second, $\delta a_2$ comprises a series of terms, each having its own scaling dimension under loop expansion~\cite{zhong24}. Correspondingly, $\delta H$ (defined as $\delta H=\delta a_2M_{\rm s}$) also contains terms with varied scaling dimensions. All these distinct dimensions emerge due to $\nu\neq1/2$ and $\delta\neq2$ despite $\beta=1$ for $d<6$~\cite{zhong24,zhong24f,zhong25}. Note that $\delta H$ is absent in Ref.~\cite{zhong24} due to the use of pre-fluctuation shift. As demonstrated below, its inclusion not only facilitates curve collapse in practice but also broadens the overlap range significantly.

Now we turn to the Ising model with the Hamiltonian 
\begin{equation}
	{\cal H} =  - \frac{J}{T}\sum\limits_{ \langle ij \rangle } {S_i S_j - \frac{H}{T}\sum\limits_i {S_i} }, \label{ham}
\end{equation}
where $S_i=\pm1$ denotes the spin on site $i$, $J$ is the coupling constant, the first summation runs over all nearest-neighbor pairs, and we have absorbed the temperature into ${\cal H}$ with the Boltzmann constant $k_{\rm B}=1$. The effective theory of the Ising model in the continuous limit is given by Eq.~(\ref{gl}), and thus they belong to the same universality class~\cite{Amit}. Here, we focus directly on the parameters in ${\cal H}$ and draw an analogy with the above theory to identify the scaling variables. Using Eq.~(\ref{th}), this yields 
\begin{equation}
\tau=J_c-J+{\cal A}+\delta J,\quad H_{\rm s}=M_{\rm s}(J_c-J+{\cal A}/3),\label{thj}
\end{equation}
where $J_c$ is the critical point of the Ising model and ${\cal A}=3a_4M_{\rm s}^2$. 

Three remarks are in order here. First, although $a_2$ is proportional to the distance to the MF critical temperature in Eq.~(\ref{f4}), $J_c$ is the exact critical point. However, $a_2+\delta a_2$ is not the exact critical point but rather a scale-dependent parameter, which is correctly represented by $J_c-J+\delta J$. Second, we have directly replaced $a_2$ with $J_c-J$. The proportionality coefficient between them can be used to rescale ${\cal A}$ in both $\tau$ and $H_{\rm s}$. Third, the temperature requires special consideration. In the time-dependent Landau-Ginzburg equation, Eqs.~(\ref{gl})--(\ref{lang2d}), a sufficiently large $T$ is essential for observing non-MF behavior~\cite{zhong24}. By contrast, in the Monte Carlo simulations used here, $T$ is built into the algorithm and combined with $J$ into a single variable. This indicates that we can have two possible schemes. The first is termed two-parameter $(J, H)$ scheme where we simply set $T=1$ in Eq.~(\ref{ham}) as the energy scale. In this way, $J$ is just the conventional inverse temperature. The second is a three-parameter $(J, H, T)$ scheme where $T$ serves as a scaled variable as in Eq.~(\ref{mha}) in the mesoscopic model~\cite{zhong24}. This additional parameter changes ${\cal H}$ by a global scale-dependent factor. Nevertheless, it can be employed to demonstrate fixed-field nucleation and growth scaling similar to that shown in Ref.~\cite{zhong24f}. As we will see below, it can also significantly extend the range of curve collapse.

To achieve curve collapse, we need to specify the considered condition. We ramp the field linearly as $H=H_{{\rm in}}+Rt$ with a constant rate $R$ and an initial value $H_{{\rm in}}<0$. Choosing appropriately the time origin such that $h=Rt$ (retaining the symbol $t$), one finds the scaling dimension of $R$ as $r\equiv[R]=z+\beta\delta/\nu$~\cite{zhongl05}. Therefore,  replacing $t$ with $R$, setting $b=R^{-1/r}$, and collecting the relevant terms in Eq.~(\ref{mha}), we reach the finite-time scaling (FTS) form~\cite{Gong,Gong1,Huang,Yuan,zhong24,zhong25} for the Ising model,
\begin{equation}
	%	\begin{split}
		M=R^{\beta/r\nu}g_2(\tau R^{ {-1/r\nu}}, hR^{-{\beta\delta/r\nu}},TR^{-[T]/r}),\label{phigr}
		%	\end{split}
\end{equation}
where $h$ is defined in Eq.~(\ref{th}), $\tau$ and $H_s$ are specified in Eq.~(\ref{thj}), and $g_2$ is yet another scaling function. FTS emerges because $b^{z}=R^{-z/r}$ represents a controllable finite timescale analogue to the finite system size in finite-size scaling~\cite{Amit}. In the two-parameter scheme, the last scaled variable in Eq.~(\ref{phigr}) is absent.

With Eq.~(\ref{phigr}), curve collapse can be achieved by fixing all scaled variables except $h$ in Eq.~(\ref{phigr}). Thus, we first run a reference curve with parameters denoted by the subscript $0$. Then, for a new $R$, we enforce the condition  $\tau_0R_0^{-1/r\nu}=\tau R^{-1/r\nu}$, resulting in,
\begin{equation}
	J = {J_c} - \left\{ {J_c} - {J_0} + {\cal A}_0\!\left[1 - ({R_0}/R)^{\rho}\right] + \delta \hat J\right\}(R_0/R)^{ - 1/r\nu } \label{je}
\end{equation}
with $\delta \hat J = \delta {J_0} - \delta J({R_0}/R)^{1/r\nu}$ and $\rho=(1-[{\cal A}]\nu)/r\nu$, where the scaling dimension of ${\cal A}$ is $[{\cal A}]=[a_4M_s^2]=(6-d)/2+\beta/\nu$~\cite{zhong24}. Within the three-parameter scheme, $T$ scales as $T=T_0(R_0/R)^{-[T]/r}$ with an arbitrary $T_0$, which is again set to 1 as the energy scale. Two parameters ${\cal A}_0$ and $\delta\hat{J}$ in Eq.~(\ref{je}) have to be estimated. We fix $\delta\hat{J}=0$ and vary ${\cal A}_0$ until the new curve aligns parallel to the reference one. This is laborious but straightforward because the curve's slope changes regularly with $J$. To collapse the two curves onto each other, we set $\delta H_0=0$ and select a $M_{\rm s0}$ to find
\begin{equation}
	H_{\rm s}=M_{\rm s0}\left\{J_c-J+{\cal A}_0(R_0/R)^{-[{\cal A}]/r}/3\right\}(R_0/R)^{-\beta/r\nu},\label{hsj}
\end{equation}
from Eq.~(\ref{thj}), where $J$ is given by Eq.~(\ref{je}). A unique $\delta H$ then collapses the rescaled curves according to Eq.~(\ref{phigr}). After determining ${\cal A}_0$, subsequent curves for other new $R$ values require only adjustments to $\delta\hat{J}$ and $\delta H$ to overlap with prior results.

It is therefore evident that the four unknown parameters ${\cal A}_0$, $\delta\hat{J}$, $M_{\rm s0}$, and $\delta H$ can be determined systematically. Nevertheless, we note that the obtained values of $\delta\hat{J}$ and $\delta H$ are based on the conditions $\delta\hat{J}=0$ (for the first non-reference curve) and $\delta H_0=0$. A different choice of the conditions gives rise to new values of $\delta\hat{J}$ and $\delta H$, which then yield a different value of $\delta J$. Therefore, although in principle $\delta J$ can be computed theoretically, its value determined numerically is only a relative one.

Moreover, $M_{\rm s0}$ can also be chosen arbitrarily even though it is theoretically fixed. Indeed, selecting a different $M_{\rm s0}$ changes the value of ${\cal A}_0$ to a new one, which can be offset by adjusting $\delta\hat{J}$ to preserve the invariance of $J$ and thus the slope. This reflects the reallocation of the ${\cal A}_0$ and $\delta\hat{J}$ values in $J$. On the other hand, the corresponding shift of the curve via Eq.~(\ref{hsj}) can be absorbed by a new $\delta H$. Therefore, different values of ${\cal A}_0$, $\delta\hat{J}$, and $\delta H$ for a different $M_{\rm s0}$ are uniquely related. In addition, as $M_{\rm s0}$ is arbitrary, $a_4$ is not necessarily directly related to $J$.
 
We simulate the Ising model on a two-dimensional square lattice with periodic boundary conditions. The system evolves through updates of randomly selected spins, with the attempted flip probability of the $i$-th spin given by $\exp(-\Delta E_i/T)/[1+\exp(-\Delta E_i/T)]$, where $\Delta E_i$ is the energy change upon flipping the spin. Time is measured in Monte Carlo steps per spin~\cite{Binderbook}. Initial conditions were verified to have negligible influence on the results for sufficiently negative $H_{\rm in}$. Averages were computed over $20,000$ independent samples. In two dimensions, ${J_c} = \ln \left(1 + \sqrt 2 \right)/2 \approx 0.4407$, $\nu=-5/2$, $\delta=-6$~\cite{Cardy85}, $\beta=1$, all are exact. $z\approx1.85$ estimated in Ref.~\cite{zhong24}. A recent three-loop calculation suggests $z\approx1.75$~\cite{Kom}, which we also employ. Accordingly, $r=z+12/5$, $\beta/r\nu=-2/5r$, $\beta\delta/r\nu=12/5r$, $[{\cal A}]=8/5$, $\rho=-2/r$, and $[T]=z-4/5$.

\begin{figure*}
%\centerline{\includegraphics{file= rgtl.eps,width=0.8\textwidth}}
\centerline{\includegraphics[width=\linewidth]{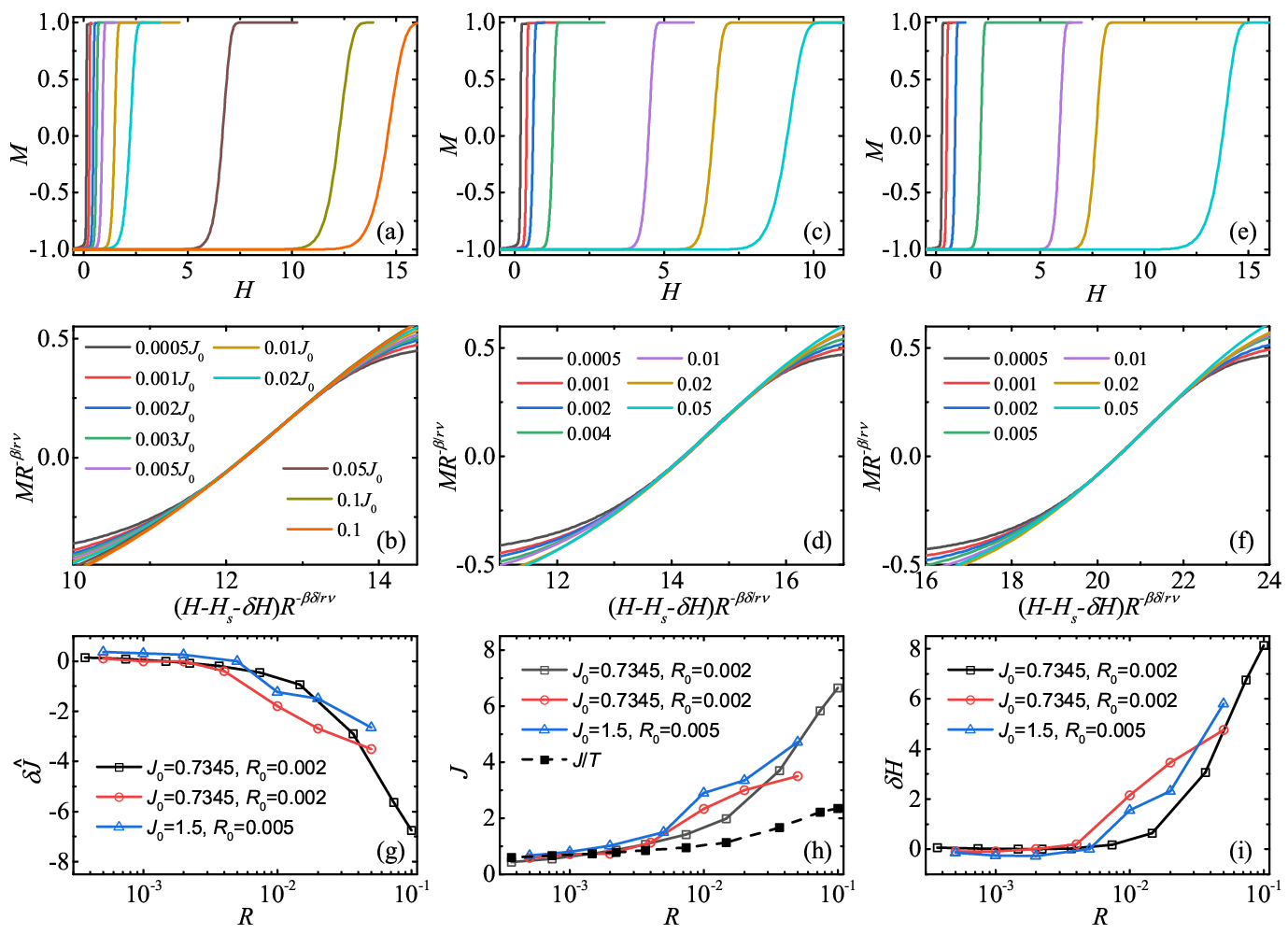}}
\caption{\label{fig} (a) $M$ versus $H$ for a series of rates $R$ listed in panel (b), obtained from Monte Carlo simulations for the three-parameter scheme. Note the $J_0$ factor of the first nine $R$ values due to a special scale choice. $J_0=0.7345$, $R_0=0.002$, $z=1.85$, ${\cal A}_0=0.35$, and $M_{\rm s0}=-0.8$. (b) Rescaling of all curves in (a) according to Eq.~(\ref{phigr}). (c) Analogous to (a) for $R$ listed in (d) but for the two-parameter scheme. $J_0=0.7345$, ${\cal A}_0=0.10$ and $M_{\rm s0}=-0.715$. (d) Analogous rescaling of all curves in (c). (e) Similar to (c) for $R$ listed in (f) but with $J_0=1.5$, $R_0=0.005$, $z=1.75$, ${\cal A}_0=0.81$, and $M_{\rm s0}=-1.5$. (f) The same rescaling of the curves in (e).  (g), (h), and (i) $R$ dependence of the used parameters $\delta\hat{J}$, $J$, and $\delta H$, respectively, for the three different choices of the parameter sets in (a), (c), and (e). The lower curve (black dashed line) in (h) represents $J/T$ of the parameter set in (a). Lines connecting symbols are only a guide to the eye.}
\end{figure*}
In Fig.~\ref{fig}(a), we present the magnetization curves for a series of $R$ spanning over two orders of magnitude for the three-parameter scheme. The corresponding $\delta\hat{J}$ and $J$ are given in Fig.~\ref{fig}(g) and~\ref{fig}(h), respectively (black curves). After rescaling and shifting with $\delta H$ depicted in Fig.~\ref{fig}(i) (also in black), the transitional parts of all these large-range curves collapse remarkably onto s single master curve, as demonstrated in Fig.~\ref{fig}(b). Such a curve collapse is far more robust than the scaling observed only at particular loci~\cite{Rao,Lo,Sengupta,He,acha,Luse,zhongpre,kim,Suen,chak,lee,zhong02,pan,zhu}, and the $R$ range is much larger than the collapse previously obtained without varying all parameters~\cite{zhong18}. In fact, the range of $R$ is unbounded: both higher and lower $R$ values yield robust collapses, in contrast to the constraints reported in Ref.~\cite{zhong24}. The reason lies in the additional parameter $\delta H$ in shifting the curves. 

Similar to previous findings in the mesoscopic models, the complete universal scaling does not collapse the entire curves due to inevitable crossovers. These crossovers occur between i) metastable states and intermediate transition states, and ii)  intermediate transition states and stable states. Both the metastable and the stable states differ fundamentally from transition states, as they are expected to be governed by distinct fixed points and exponents rather than those associated with the cubic universality class. The interplay between these fixed points generates the upper and lower crossover regions, which link the intermediate transition region. Both the intermediate region and the crossover regions exhibit $R$-dependence, as seen in Figs.~\ref{fig}(b),~\ref{fig}(d), and~\ref{fig}(f).

In Fig.~\ref{fig}(c) and~\ref{fig}(d), we show similar results of the two-parameter scheme using the same $J_0$ and $R_0$. A different ${\cal A}_0$ is required and a new $M_{\rm s0}$ is selected. Nevertheless, good curve collapse is evident. In Fig.~\ref{fig}(e) and~\ref{fig}(f), we further demonstrate the two-parameter scheme for another set of $J_0$ and $R_0$ with a new $z=1.75$, distinct from the above two cases. One sees that the collapse exhibits comparative quality, although this implies that $z$ cannot be uniquely determined via optimal collapse criteria~\cite{zhong24}. The two distinct sets of ${\cal A}_0$ and $M_{\rm s0}$ yield $a_{40}$ values of $0.065$ and $0.12$, respectively. Their ratio is close to, yet does not match, that of the corresponding $J$ values. As noted earlier, this mismatch results from the arbitrariness in choosing $M_{\rm s0}$. Nevertheless, all sets of curves exhibit robust collapse as demonstrated in Figs.~\ref{fig}(b),~\ref{fig}(d), and~\ref{fig}(f), unambiguously confirming the underlying scale invariance in FOPTs. 

We note that the key difference between the three- and the two-parameter scheme is that $R$ cannot be too large in the latter. As illustrated in Fig.~\ref{fig}(h), for large $R$, $J$ of the two-parameter scheme (red and blue curves) significantly exceeds the actual interaction $J/T$ (black dashed curve) of the three-parameter scheme. Consequently, the interaction coefficient is so large that the curve's slope changes little. For comparison, $T=(R_0/R)^{-[T]/r}>1$ for $R>R_0$ suppresses $J$ in the three-parameter scheme. Note that for small $R$, $J$ is naturally small, rendering the $T$ factor unimportant.

The results for $\delta\hat{J}$, $J$, and $\delta{H}$ depicted in
Figs.~\ref{fig}(g)--\ref{fig}(i) demonstrate that these quantities do not exhibit simple power-law relations with $R$. This absence of scaling arises because $\delta\hat{J}$ and $\delta{H}$ both are composed of multiple terms with distinct exponents, as noted earlier. The different variations of these quantities with $R$ stem from the nonlinear relationship with $M_{\rm s0}$.

%\section{Conclusion}
We have successfully applied the mesoscopic theory to the microscopic Ising model, demonstrating complete universal scaling in its FOPTs below the critical temperature over a broad range of the ramping rates. This is accomplished through a direct mapping between the parameters of the microscopic model and the mesoscopic theory. A systematic determination of the four unknown parameters, ${\cal A}_0$, $\delta\hat{J}$, $\delta H$, and  $M_{\rm s0}$, with $M_{\rm s0}$ being arbitrary numerically, along with two schemes for the distinct roles of the temperature, has been developed and validated. A post-fluctuation shift embodied in the parameter $\delta H$ significantly helps curve collapse. Additionally, different dynamic exponents have been found to work equally well. Our findings further confirm that scaling and universality, previously recognized in conventional continuous phase transitions, also emerge in FOPTs. Furthermore, this work provides a framework for extending the theory to other systems exhibiting field-driven FOPTs, including experimentally accessible artificial systems such as cold atoms. For real systems, in which interactions are not readily adjustable, experimental scaling may possibly be approximate.  

\begin{acknowledgments}
This work was supported by National Natural Science Foundation of China (Grant No. 12175316).
\end{acknowledgments}

\end{document}